\documentclass[preprint]{elsarticle}
\usepackage{dcolumn}
\usepackage{graphicx}
\usepackage{float}
\usepackage{physics}
\usepackage{lineno}
\usepackage{geometry}
\usepackage[colorlinks=true,allcolors=blue]{hyperref}

\journal{Chinese Physics C}

\begin{document}

\begin{frontmatter}

\title{Estimation of the chiral magnetic effect considering the magnetic field response of the QGP medium}

\author[ctgu,moe,wuhan]{Sheng-Qin Feng\corref{correfauthor}}
\cortext[correfauthor]{Corresponding author}
\ead{fengsq@ctgu.edu.cn}

\author[ctgu]{Xin Ai}

\author[ctgu]{Lei Pei}

\author[ctgu]{Fei Sun}

\author[ctgu]{Yang Zhong}

\author[moe]{Zhong-Bao Yin}

\address[ctgu]{College of Science, China Three Gorges University, Yichang 443002, China}
\address[moe]{Key Laboratory of Quark and Lepton Physics (MOE) and Institute of Particle Physics,\\
Central China Normal University, Wuhan 430079, China}
\address[wuhan]{School of Physics and Technology, Wuhan University, Wuhan 430072, China}

\begin{abstract}
The magnetic field plays a major role in the searching of the chiral magnetic effect in relativistic heavy-ion collisions. If the lifetime of the magnetic field is too short, as expected by simulations of the field in the vacuum, the chiral magnetic effect will be largely suppressed. However, the lifetime of the magnetic field will become longer when the QGP medium response is considered. We give an estimate of the effect, especially considering the magnetic field response of the QGP medium, and compare it with the experimental results of background-subtracted correlator $H$ at RHIC and LHC energies. The results show that our method explains better for the experimental results at the top  RHIC energy than that of the LHC energy.
\end{abstract}


\end{frontmatter}


\section{Introduction}\label{intro}
The interplay of quantum anomaly and magnetic field leads to a lot of macroscopic quantum phenomena in relativistic heavy-ion collisions. The most important one that we discuss here is the chiral magnetic effect (CME). The CME is the separation of electric charge along the magnetic field in the presence of chirality imbalance\cite{Kharzeev:2007jp,Warringa:2008kv,Fukushima:2008xe}. It has already been observed in condensed matter systems\cite{Li:2014bha}.

The question is whether the CME exists in relativistic heavy-ion collisions. The answer seems to be yes. Two necessary conditions, chirality imbalance, and magnetic field may be met in QGP produced in relativistic heavy-ion collisions. Firstly, quantum chromodynamics (QCD) which describes the behavior of the QGP permits topological charge changing transition that can induce chirality imbalance\cite{Kharzeev:2007jp}. Secondly, enormous magnetic field can be produced in non-central relativistic heavy-ion collisions due to charged nucleus moving at speed close to the speed of light\cite{Skokov:2009qp,Voronyuk:2011jd,Bzdak:2011yy,Deng:2012pc,Tuchin:2013apa,Mo:2013qya}. Therefore, the CME is very likely to exist in relativistic heavy-ion collisions.

Over the past few years, much effort has been given to the search of the experimental evidence of the CME in relativistic heavy-ion collisions. Several collaborations at the BNL Relativistic Heavy Ion Collider (RHIC), and the CERN Large Hadron Collider (LHC), including STAR\cite{Abelev2009ac, Abelev:2009ad, Voloshin:2008jx, PhysRevC.88.064911, Wang:2012qs, Adamczyk:2014mzf, Adamczyk:2013kcb}, PHENIX\cite{Ajitanand:2010}, and ALICE\cite{Christakoglou:2011uqg,PhysRevLett.110.012301} have studied this; for recent reviews see Refs.~\cite{Wang:2016mkm}.
At first glance, it seems easy to detect the CME experimentally. In fact, this is not the case. Firstly, one cannot identify the charge asymmetry in an individual event as the sign of the CME. This is due to the fact that statistical fluctuations $\sim \sqrt{N}$ is much larger than the expected charge asymmetry induced by the CME, where $N$ is the charged-particle multiplicity of produced particles\cite{Kharzeev:2007jp}. However, if one takes an average over many events directly, the contributions of the CME will also be canceled out, since the right-handed and left-handed chirality is produced with equal probability.

One proposed to measure the charge separation fluctuations perpendicular to the reaction plane by a three-point correlator, $\gamma \equiv \langle \langle \cos(\phi_\alpha + \phi_\beta - 2 \Psi_\text{RP}) \rangle \rangle$, where the averaging is done over all particles in an event and over all events\cite{Voloshin:2004vk,Kharzeev:2007jp}. This correlator will remove the multiplicity fluctuations while keeping the contributions from the CME. The $\gamma$ correlator was first measured by the STAR Collaboration for Au+Au and Cu+Cu collisions at $62.4$ and $200\,\mathrm{GeV}$\cite{Abelev2009ac,Abelev:2009ad}. All the results have been found to be qualitatively consistent with the theoretical expectation of the CME. Similar results have also been observed by the ALICE Collaboration for $2.76\,\mathrm{TeV}$ Pb+Pb collisions\cite{PhysRevLett.110.012301}.

Unfortunately, the $\gamma$ correlator still contains some background contributions not related to the CME\cite{Schlichting:2010qia,Pratt:2010zn,Bzdak:2012ia}. These background contributions are mainly from the elliptic flow in combination with two-particle correlations. To solve this problem, one introduced the two-particle correlator, $\delta \equiv \langle \cos(\phi_\alpha - \phi_\beta) \rangle$. Similar to the $\gamma$ correlator, the $\delta$ also contains the contributions from the CME and the backgrounds, but it is dominated by backgrounds. It is suggestive to express the $\gamma$ and the $\delta$ in the following ways\cite{Bzdak:2012ia, Liao:2015Pr}:
\begin{align}
\gamma &= \kappa v_2 B - H, \label{eq:gamma} \\
\delta &= B + H, \label{eq:delta}
\end{align}
where $H$ and $B$ are the CME and background contributions, respectively. The background-subtracted correlator, $H$, can be obtained by solving Eq.~(\ref{eq:gamma}) and (\ref{eq:delta}):
\begin{equation}
H^\kappa = \frac{\kappa v_2 \delta - \gamma}{1 + \kappa v_2}.
\end{equation}

The coefficient $\kappa$ is close to but deviates from unity owing to the finite detector acceptance and theoretical uncertainties\cite{Bzdak:2012ia}. The $\delta$ correlators for $200\,\mathrm{GeV}$ Au+Au collisions and $2.76\,\mathrm{TeV}$ Pb-Pb collisions have been measured by STAR\cite{Abelev:2009ad} and ALICE\cite{PhysRevLett.110.012301}, respectively. The correlator $H_\text{SS} - H_\text{OS}$ has been measured by STAR for Au+Au collisions at $\sqrt{s_\text{NN}} = 7.7$--$62.4\,\mathrm{GeV}$\cite{Adamczyk:2014mzf}. The results show that there is a clear charge-separation effect at $\sqrt{s_\text{NN}} = 19.6$--$200\,\mathrm{GeV}$ for mid-peripheral ($30$--$80\%$ centrality) collisions. It is again in line with the expectations of the CME.

To better explain the experimental results, a quantitative estimation of the CME is needed.
In Ref.~\cite{Kharzeev:2007jp}, Kharzeev, Mclerran, and Warringa (KMW) developed a quantitative model to estimate the CME induced charge separation.

One of the main issue in estimating the CME is the time evolution of the magnetic field in relativistic heavy-ion collisions.
This issue has been studied by many works of literature\cite{Kharzeev:2007jp,Skokov:2009qp,Voronyuk:2011jd,Bzdak:2011yy,Deng:2012pc,Tuchin:2013apa,Mo:2013qya}. The numerical calculations carried out by these works of literature show that an enormous magnetic field ($B \sim 10^{15}\,\mathrm{T}$) can be found at the very beginning of the collisions.
However, according to these studies, the strength of the magnetic field decreases rapidly with time.
It is a challenge for the manifestation of the CME in relativistic heavy-ion collisions. If the lifetime of the magnetic field is too short, the imprint of the CME might be negligible.
Nevertheless, one proposed that these estimations of the magnetic field are valid only at the early stage of the collision. At a later time, the magnetic response from the QGP medium becomes increasingly important\cite{Tuchin:2010vs,Tuchin:2010vs2,Deng:2012pc,Tuchin:2013apa,McLerran:2013hla,Tuchin:2013ie,Zakharov:2014dia,Tuchin:2014iua,Tuchin:2015oka}, and the magnetic field will maintain a much longer time than in the vacuum.

This work aims to give an estimation of the CME, especially considering the magnetic response of the QGP medium, and then compare it with the experimental results of background-subtracted correlator $H$.

This paper is organized as follows. We give an introduction of the KMW model in Sec.~2. The time evolution of the magnetic field in relativistic heavy-ion collisions is discussed in Sec.~3. In Sec.~4, we present our computation results. A summary is given in Sec.~5.

\section{The KMW model for the CME} \label{KMWmodel}

In this section, we will briefly introduce the KMW model for estimating the CME in relativistic heavy-ion collisions.

All gauge field configurations which have finite action can be categorized into topologically distinct classes labeled by the winding number $Q_\text{w}$. Configurations with non-zero $Q_\text{w}$ can induce chirality imbalance through the axial anomaly. If initially there are an equal number of right-handed and left-handed fermions, i.e., $N_R = N_L$, at $t = \infty$ we have
\begin{equation}
(N_L - N_R)_{t=\infty} = 2N_f Q_\text{w}.
\end{equation}

The classical vacuum of QCD is degenerate, and the winding number $n_\text{w}$ can characterize the different classical vacua. It can be showed that if a gauge field configuration with non-zero $Q_\text{w}$ goes to a pure gauge at infinity, it induces a transition from one classical vacuum to another.

The transition can be achieved through instanton\cite{Diakonov:2002fq,Schafer:1995pz} or sphaleron\cite{Arnold:1987zg,Fukugita:1990gb}. The instanton corresponds to quantum tunneling through the energy barrier between different QCD vacuum which is highly suppressed. However, the sphaleron corresponds to go over the barrier, and its transition rate can be very high at high temperature which happens to be the situation of the QGP. Thus it provides the chance to induce chirality.

The transition rate for the QCD has been estimated in Ref.~\cite{Kharzeev:2007jp} as follows:
\begin{equation} \label{tranRate}
\frac{\dd{N^\pm_\text{t}}}{\dd[3]{x}\dd{t}} \equiv \Gamma^\pm \sim 192.8 \alpha^5_S T^4,
\end{equation}
where the superscript $\pm$ denotes the transitions with $Q_\text{w} = \pm 1$. The total rate of transition is the sum of the rates of the lowering and rising transition,
\begin{equation}
\frac{\dd{N_\text{t}}}{\dd[3]{x}\dd{t}} = \sum_\pm \frac{\dd{N_\text{t}^\pm}}{\dd[3]{x}\dd{t}}.
\end{equation}

In the case of a sufficiently large magnetic field, the charge separation perpendicular to the magnetic field induced by a configuration with winding number $Q_\text{w}$ is as follows
\begin{equation}
Q = 2 Q_\text{w} \sum_f |q_f|,
\end{equation}
where $q_f$ is the charge in units of $e$ of a quark with flavor $f$. For a moderate magnetic field, the estimation given by Ref.~\cite{Kharzeev:2007jp} is
\begin{equation}
Q \approx 2 Q_\text{w} \sum_f |q_f| \gamma(2|q_f \Phi|),
\end{equation}
where
\begin{equation}
\gamma(x) = \begin{cases}
x, & \text{for } x \leq 1, \\
1, & \text{for } x \geq 1,
\end{cases}
\end{equation}
and $\Phi = eB\rho^2$ is the flux through a configuration of size $\rho$ with non-zero $Q_\text{w}$.

Now we consider the situation in relativistic heavy-ion collisions. We use the same symbols defined in Ref.~\cite{Kharzeev:2007jp}. $N^\pm_a$ and $N^\pm_b$ denote the total positive/negative charge in units of $e$ above (a) and below (b) the reaction plane respectively; $\Delta_\pm$ is the difference between in charge on each side of the reaction plane $\Delta_\pm = N^\pm_a - N^\pm_b$.

When there is a transition from one vacuum to another, a charge difference will be created locally. However, the quarks may encounter many interactions in the QGP, and this will suppress the degree of the final observed charge separation. In considering this, the screening suppression functions $\xi_\pm(x_\perp)$ are introduced in Ref.~\cite{Kharzeev:2007jp}. The expression is as follows
\begin{equation}
\xi_\pm(x_\perp) = \exp(-|y_\pm(x) - y|/\lambda),
\end{equation}
where $\lambda$ is the screening length and $y_\pm(x)$ is the upper and lower $y$ coordinate of the overlap region. The expectation value of the change of the $\Delta_+$ and $\Delta_-$ due to a transition is either positive or negative with equal probability and given by
\begin{equation}
\pm \sum_f |q_f| \gamma(2|q_f \Phi|) \xi_\pm(x_\perp).
\end{equation}
Here only the most probable transitions have been considered, namely $Q_\text{w} = \pm 1$.

By assuming that all transitions happen independently from each other, one can compute the variation of $\Delta^\pm$:
\begin{align} \label{delta2}
\langle \Delta^2_\pm \rangle &= \frac{1}{2}\int_{t_i}^{t_f} \dd{t} \int_V \dd[3]{x} \int \dd{\rho} \frac{\dd{N_t}}{\dd[3]{x}\dd{t}\dd{\rho}} \\*
 &\times [\xi_-(x_\perp)^2 + \xi_+(x_\perp)^2] \Bigl[ \sum_f |q_f| \gamma(2|q_f eB|\rho^2) \Bigr]^2, \nonumber
\end{align}
and $\langle \Delta_+ \Delta_- \rangle$ can also be calculated:
\begin{align}\label{delta+-}
\langle \Delta_+ \Delta_- \rangle &= - \int_{t_i}^{t_f} \dd{t} \int_V \dd[3]{x} \int \dd{\rho} \frac{\dd{N_t}}{\dd[3]{x}\dd{t}\dd{\rho}} \\*
&\times \xi_-(x_\perp) \xi_+(x_\perp) \Bigl[ \sum_f |q_f| \gamma(2|q_f eB|\rho^2) \Bigr]^2. \nonumber
\end{align}

In Ref.~\cite{Kharzeev:2007jp}, the Eq.~(\ref{delta2}) and (\ref{delta+-}) have been rewritten for small magnetic fields ($2|q_feB|<1/\rho^2$) using the Eq.~(\ref{tranRate}) for transition rate and the fact that $\rho \sim (\Gamma^\pm / \alpha_S)^{-1/4} \sim 1 / (\alpha_S T)$. They are given as follows
\begin{align} \label{eq:deltapm}
\dv{\langle \Delta^2_\pm \rangle}{\eta} &= 2\kappa \alpha_S \Bigl[ \sum_f q_f^2 \Bigr]^2 \int_{V_\perp} \dd[2]{x_\perp} \\*
&\times [\xi_-(x_\perp)^2 + \xi_+(x_\perp)^2] \int_{\tau_i}^{\tau_f} \dd{\tau} \tau [eB(\tau, \eta, x_\perp)]^2, \nonumber \\ \label{eq:deltapm2}
\dv{\langle \Delta_+ \Delta_- \rangle}{\eta} &= -4\kappa \alpha_S \Bigl[ \sum_f q_f^2 \Bigr]^2 \int_{V_\perp} \dd[2]{x_\perp} \\*
&\times \xi_+(x_\perp)\xi_-(x_\perp) \int_{\tau_i}^{\tau_f} \dd{\tau} \tau [eB(\tau, \eta, x_\perp)]^2, \nonumber
\end{align}
where the proper time $\tau = (t^2 - z^2)^{1/2}$ and the space-time rapidity $\eta = \frac{1}{2} \log[(t+z)/(t-z)]$. The volume integral is over the overlap region $V_\perp$ in the transverse plane. The assumption here is that the magnetic field does not change the transition rate dramatically. There is also a constant $\kappa$ for which the order of magnitude should be one, but with large uncertainties\cite{Kharzeev:2007jp}.

In Ref.~\cite{Kharzeev:2007jp}, they connect the $\langle \Delta^2_\pm \rangle$ and $\langle \Delta_+ \Delta_- \rangle$ to correlators $a_{++}$($a_{+-}$) by expressions
\begin{align} \label{eq:app}
a_{++} &= a_{--} = \frac{1}{N_+^2} \frac{\pi^2}{16} \langle \Delta_\pm^2 \rangle, \\ \label{eq:apm}
a_{+-} &= \frac{1}{N_+ N_-} \frac{\pi^2}{16} \langle \Delta_+ \Delta_-  \rangle,
\end{align}
where $N_\pm$ is the total number of positively or negatively charged particles in the corresponding $\eta$ interval. The correlator $a_{++}$($a_{+-}$) is the same as the $\gamma$ correlator, except a sign difference. However, in this model the $v_2$-related backgrounds are completely ignored; thus we should compare the model calculated correlators $a_{++}$($a_{+-}$) with the background-subtracted correlator $H_\text{SS}$($H_\text{OS}$). Because $H$ also has a sign difference with $\gamma$ as shown in Eq.~(\ref{eq:gamma}), so there is no sign difference between the $a_{++}$($a_{+-}$) and $H_\text{SS}$($H_\text{OS}$).

\section{Magnetic field in relativistic heavy-ion collisions}\label{eBinHIC}
We will discuss the magnetic field in relativistic heavy-ion collisions in this section. Ref.~\cite{Kharzeev:2007jp} gave a calculation of the magnetic field in relativistic heavy-ion collisions. The calculation was done by an analytic model with the assumption that the nucleon density is uniform in rest frame. On the basis of it, Ref.~\cite{Mo:2013qya} improved the calculation by using the Woods-Saxon nucleon distribution. There are also many other calculations using different methods\cite{Skokov:2009qp,Voronyuk:2011jd,Bzdak:2011yy,Deng:2012pc,Tuchin:2013apa}.

However, most of these calculations did not consider the magnetic response of the QGP medium which may notably influence the time evolution of the magnetic field. Tuchin first analyzed it in Ref.~\cite{Tuchin:2010vs,Tuchin:2010vs2}, and he concluded that the magnetic field is almost constant during the entire plasma lifetime due to high electric conductivity.
Later, it was quantitatively studied by many works of literature\cite{Deng:2012pc,Tuchin:2013apa,McLerran:2013hla,Tuchin:2013ie,Zakharov:2014dia,Tuchin:2014iua,Tuchin:2015oka}. To explore this problem, one needs considering the electric conductivity $\sigma$ and chiral magnetic conductivity $\sigma_\chi$ which is induced by the CME.  In Ref.~\cite{McLerran:2013hla}, they found the effects of finite $\sigma_\chi$ are not important for the top RHIC and LHC energies. Therefore, we are not considering the effects of chiral magnetic conductivity in this paper. For electric conductivity $\sigma$, there are a lot of theoretical uncertainties\cite{Arnold:2003zc, Gupta:2003zh, Ding:2010ga, Francis:2011bt}.

For computation simplicity, we adopt the most optimistic situation proposed in Ref.~\cite{Deng:2012pc}, namely assuming the electric conductivity $\sigma$ is large enough that we can take the QGP as an ideally conducting plasma. Under this assumption, one gets the following equations from Maxwell's equations:
\begin{align}
\frac{\partial \vb{B}}{\partial t} &= \nabla \times (\vb{v} \times \vb{B}),\label{eq:pBpt}  \\
\vb{E} &= - \vb{v} \times \vb{B}, \label{eq:E}
\end{align}
where the $\vb{v}$ is the flow velocity of QGP.

To solve the above equations, one needs to know the evolution of $\vb{v}$. In Ref.~\cite{Deng:2012pc}, they assumed the Bjorken picture for the longitudinal expansion,
\begin{equation}
v_z = \frac{z}{t}.
\end{equation}
For transverse expansion, they applied a linearized ideal hydrodynamic equation proposed by Ref.~\cite{Ollitrault:2008zz}, and got the following solution
\begin{align}
v_x &= \frac{c_s^2}{a_x^2} xt, \\
v_y &= \frac{c_s^2}{a_y^2} yt,
\end{align}
where $c_s$ is the speed of sound, and $a_{x,y}$ is the root-mean-square of the transverse entropy distribution. Here, we take $a_x \sim a_y \sim 3$ and $c_s^2 \sim 1/3$.

Substituting the velocity into the Eq.~(\ref{eq:pBpt})-(\ref{eq:E}), one can solve $\vb{B}(t)$ for a given initial condition $\vb{B}^0(\vb{r}) = \vb{B}(t=t_0, \vb{r})$ where $t_0$ is the formation time of the QGP. Here, we only consider the $y$ component of the magnetic field at the center of the collision region, and one gets the following solution
\begin{equation}
B_y(t,\vb{0}) = \frac{t_0}{t} e^{-\frac{c_s^2}{2a_x^2}(t^2 - t_0^2)} B_y^0(\vb{0}). \label{eq:eBinQGP}
\end{equation}

To get the time evolution of magnetic field from Eq.~(\ref{eq:eBinQGP}), we must know the formation time $t_0$ of the QGP and the initial magnetic field at that time, namely $B_y^0(\vb{0})$. For initial magnetic field, we use the method in Ref.~\cite{Mo:2013qya} which doesn't consider the QGP medium response. For the formation time of QGP, the following approximation formula has been used:
\begin{equation}
t_0 \sim 1 / Q_{s},
\end{equation}
where $Q_s$ is the saturation momentum.

The value of saturation momentum $Q_s$ for Au-Au collisions at $\sqrt{s} = 130\,\mathrm{GeV}$ is provided by Ref.~\cite{Kharzeev:2000ph}. We use the following formula for estimating the energy and nuclear dependence of the saturation momentum\cite{Kowalski:2007rw}:
\begin{equation}
Q_s^2 \sim A^{1/3} x^{-0.3}, \label{eq:Qs}
\end{equation}
where Bjorken $x = Q_s/\sqrt{s}$. Then, the saturation momentum for collisions with the different nucleus and center-of-mass energy can be calculated by the results of Ref.~\cite{Kharzeev:2000ph} using Eq.~(\ref{eq:Qs}).

\begin{figure}
\centering
\includegraphics[width=8.5cm]{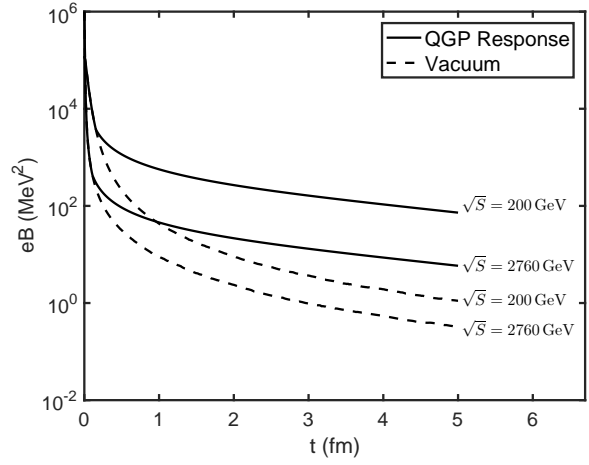}
\caption{\label{fig1}The time evolution of magnetic field for Au-Au collisions with $b = 8\,\mathrm{fm}$ at $\sqrt{s}=200\,\mathrm{GeV}$ and Pb-Pb collisions with $b = 8\,\mathrm{fm}$ at $\sqrt{s} = 2760\,\mathrm{GeV}$. The solid line and dashed line represent with and without considering QGP medium, respectively.}
\end{figure}

The centrality dependence of $Q_s^2$, $t_0$, and $eB_y^0$ for Au-Au collisions at RHIC energy and Pb-Pb collisions at LHC energy have been given in Tab.~\ref{tab:RHICdata}--\ref{tab:LHCdata} where the average impact parameter $b$ is inferred from Ref.~\cite{Ray:2007av,Abelev:2013qoq}. The time evolution of magnetic field is plotted in Fig.~\ref{fig1}, and the magnetic field in vacuum is also added for comparison.

\begin{table}
\centering
\caption{\label{tab:RHICdata}Centrality dependence  of $Q_s^2$, $t_0$ and $eB_y^0$ for Au-Au collisions at $\sqrt{s} = 200\,\mathrm{GeV}$.}
\begin{tabular}{lllll}
\hline
  Centrality & $b$ & $Q_s^2$ & $t_0$ & $eB_y^0$ \\
             & ($\mathrm{fm}$)  &  ($\mathrm{GeV}^2$) & ($\mathrm{fm}$) & ($\mathrm{MeV}^2$) \\
\hline
0--5\% & 2.21 & 2.25 & 0.132 & 2161.4 \\
5--10\% & 4.03 & 2.15 & 0.135 & 3382.9 \\
10--20\% & 5.70 & 1.99 & 0.140 & 3942.6 \\
20--30\% & 7.37 & 1.75 & 0.149 & 3909.5 \\
30--40\% & 8.73 & 1.50 & 0.161 & 3447.2 \\
40--50\% & 9.90 & 1.22 & 0.179 & 2771.0 \\
50--60\% & 11.00 & 0.92 & 0.205 & 2001.3 \\
  \hline
\end{tabular}
\end{table}

\begin{table}
\centering
\caption{\label{tab:LHCdata}Centrality dependence of $Q_s^2$, $t_0$ and $eB_y^0$ for Pb-Pb collisions at $\sqrt{s} = 2760\,\mathrm{GeV}$.}
\begin{tabular}{lllll}
\hline
  Centrality & $b$ & $Q_s^2$ & $t_0$ & $eB_y^0$ \\
             & ($\mathrm{fm}$)  &  ($\mathrm{GeV}^2$) & ($\mathrm{fm}$) & ($\mathrm{MeV}^2$) \\
\hline
0--5\% & 2.43 & 4.52 & 0.093 & 700.8 \\
5--10\% & 4.31 & 4.33 & 0.095 & 782.8 \\
10--20\% & 6.05 & 4.01 & 0.099 & 673.4 \\
20--30\% & 7.81 & 3.53 & 0.105 & 469.3 \\
30--40\% & 9.23 & 3.01 & 0.114 & 287.0 \\
40--50\% & 10.47 & 2.45 & 0.126 & 151.1 \\
50--60\% & 11.58 & 1.86 & 0.145 & 64.5 \\
  \hline
\end{tabular}
\end{table}

\begin{figure}
\centering
\includegraphics[width=8.5cm]{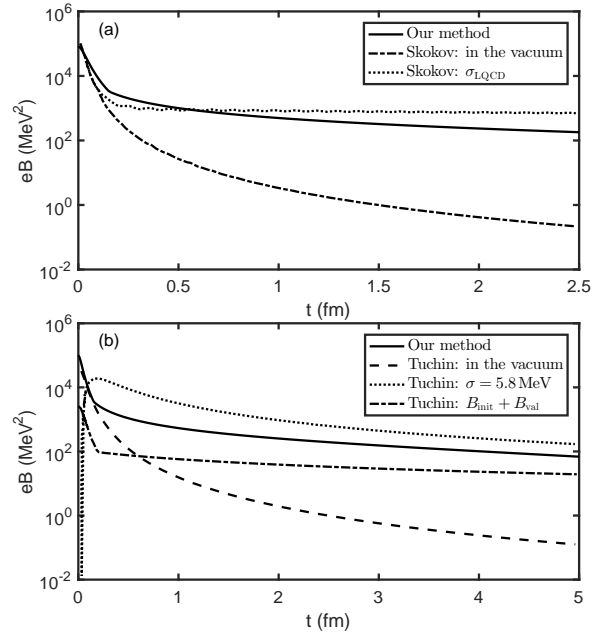}
\caption{\label{fig2}The comparisons among our results with Skokov's(a) and Tuchin's(b).}
\end{figure}

We also compare our results with the results of Skokov\cite{McLerran:2013hla} and Tuchin\cite{Tuchin:2013ie,Tuchin:2015oka}, which are plotted in Fig.~\ref{fig2}. In Fig.~\ref{fig2}(a), we compare with Skokov's results for Au-Au collisions at $\sqrt{s} = 200\,\mathrm{GeV}$ and $b = 6\,\mathrm{fm}$. The solid line represents our method; the dash-dotted line represents the Skokov's result in the vacuum; the dotted line represents the Skokov's result with the conductivity set by lattice QCD calculation.

By comparison, we can find that Skokov's magnetic field drops more rapidly than our's in the beginning, and it goes down more slowly at the later time. The possible causes of the difference are as follows: the different settings in electric conductivity $\sigma$, the neglection of the influences of flow velocity $\vb{v}$ and the QGP formation time $t_0$ by Skokov.

In Fig.~\ref{fig2}(b), we compare with Tuchin's results for Au-Au collisions at $\sqrt{s} = 200\,\mathrm{GeV}$ and $b = 7\,\mathrm{fm}$. The solid line also represents our method; the dashed line represents the Tuchin's result in the vacuum. The dotted line represents the Tuchin's result by setting electric conductivity $\sigma = 5.8\,\mathrm{MeV}$\cite{Tuchin:2013ie}. However, Ref.~\cite{Tuchin:2013ie} did not consider the contributions from the initial magnetic field, and also ignored the QGP formation time $t_0$. This explains why the magnetic field increases rapidly from zero at the beginning. The result of  considering the initial magnetic field\cite{Tuchin:2015oka} is plotted by the dash-dotted line in Fig.~\ref{fig2}(b) with the QGP formation time $t_{0} = 0.2\,\mathrm{fm}$. Ref.~\cite{Tuchin:2015oka} simplified relativistic heavy-ion collision as two counter-propagating charges which may be the main reason for the difference in magnitude with our result. Nevertheless, the overall trend of our result is very similar with that of the Ref.~\cite{Tuchin:2015oka}.

\section{Computation results}\label{compuResult}
In this section, we are going to give an estimation of the CME in relativistic heavy-ion collisions using the KMW model introduced in Sec.~2.

\begin{figure}
\centering
\includegraphics[width=8.5cm]{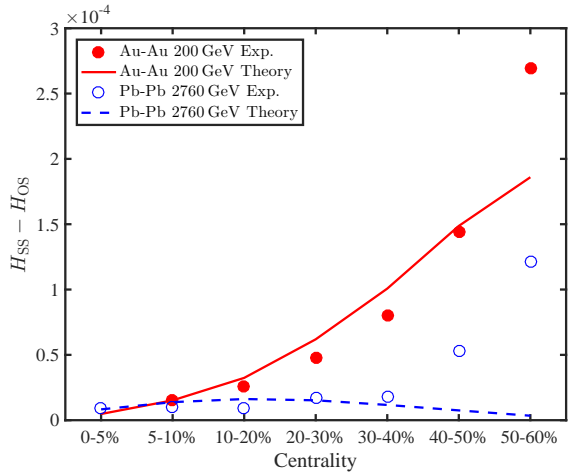}
\caption{\label{fig3}The comparison between the centrality dependence of the KMW model estimated $a_{++}-a_{+-}$ with background-subtracted experiment observable $H_\text{SS} - H_\text{OS}$.}
\end{figure}

We use the Eqs.~(\ref{eq:deltapm})--(\ref{eq:apm}) to determine the centrality dependence of correlator $a_{++}$($a_{+-}$). The time evolution of magnetic field has been discussed in Sec.~3. For correspondence between impact parameter and centrality, we refer to Refs.~\cite{Ray:2007av,Abelev:2013qoq}.
The number of charged particles $N_\pm$ is obtained from Refs.~\cite{Ray:2007av,Aamodt:2011cdo}. As explained in Sec.~1, the KMW model does not consider the contributions from the background, so we compare our results with the experimental results of background-subtracted correlator $H$. The undetermined parameters $\chi$ and $\lambda$ is fixed by fitting the experiment observable $H_\text{SS} - H_\text{OS}$. The results for Au-Au collisions at $\sqrt{s} = 200\,\mathrm{GeV}$ and Pb-Pb collisions at $\sqrt{s} = 2760\,\mathrm{GeV}$ are plotted in Fig.~\ref{fig3}.

It can be seen from Fig.~\ref{fig3} that the model explains the experimental data better at RHIC than LHC energy. For Au-Au collisions at $\sqrt{s} = 200\,\mathrm{GeV}$, the general trend is consistent with the experiment, but it deviates from the experiment at periphery collision. This may due to the hard-sphere approximation which is used in determining the overlap region $V_\perp$ in Eq.~(\ref{eq:deltapm}) and Eq.~(\ref{eq:deltapm2}).

For Pb-Pb collisions at $\sqrt{s} = 2760\,\mathrm{GeV}$, it rises with centrality goes up (more periphery) at central collisions and then falls at periphery collisions. This trend is completely different from the experimental data. The reasons for its fall at periphery collisions are as follows. In general, the magnetic field in vacuum increases with the increases of impact parameter $b$. However, the magnetic field considering QGP medium has a strong dependence on QGP formation time $t_0$ at high energy. As we can see from Fig.~\ref{fig1}, the magnetic field drops more quickly at high energy. Therefore, a slight change in $t_0$ will greatly influence the magnetic field and then the CME. Besides, from Tab.~\ref{tab:RHICdata}--\ref{tab:LHCdata} we know that $t_0$ becomes larger at periphery collisions. The combination of these leads to the falls of correlator $a_{++} - a_{+-}$ in periphery collisions. This effect also exists in Au-Au collisions at $\sqrt{s} = 200\,\mathrm{GeV}$, but it is weaker at low energy.

This discrepancy between the theory and experiment at high energy reflects the shortcomings of our model. This may be because we only consider the magnetic field at the origin, namely $B_y(t,\vb{0})$, for simplicity. It is appropriate when the magnetic field is homogeneous. At high energy, however, the magnetic field may be highly inhomogeneous in beam direction. Therefore, only considering the magnetic field at the origin will largely underestimate the overall effects. This problem should be further studied at later works.

\begin{figure}
\centering
\includegraphics[width=8.5cm]{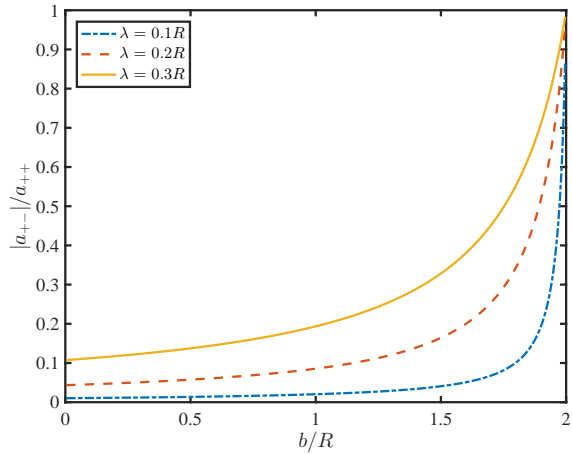}
\caption{\label{fig4}The results of the correlator $|a_{+-}|/a_{++}$ as a function of $b/R$ with different screening length $\lambda$ for Au-Au collisions at $\sqrt{s} = 200\,\mathrm{GeV}$.}
\end{figure}

Generally, the correlator $a_{+-}$ is less than $a_{++}$ because of the screening effect. We plot the centrality dependence of $|a_{+-}|/a_{++}$ with different screening length for Au-Au collisions at $\sqrt{s} = 200\,\mathrm{GeV}$ in Fig.~\ref{fig4}. The results are similar to the results in Ref.~\cite{Kharzeev:2007jp}: the correlator $|a_{+-}|/a_{++}$ increases as impact parameter $b$ increases. It is because that the system size is small when the impact parameter $b$ is large, and the smaller the system size, the weaker the screening effect. Note that the weaker the screening effect, the bigger the correlator $|a_{+-}|/a_{++}$ and when there is no screening effect, the correlator $|a_{+-}|/a_{++}$ should equal to $1$. This also explains why the correlator $|a_{+-}|/a_{++}$ increases as screening length $\lambda$ increases.

\begin{figure}
\centering
\includegraphics[width=8.5cm]{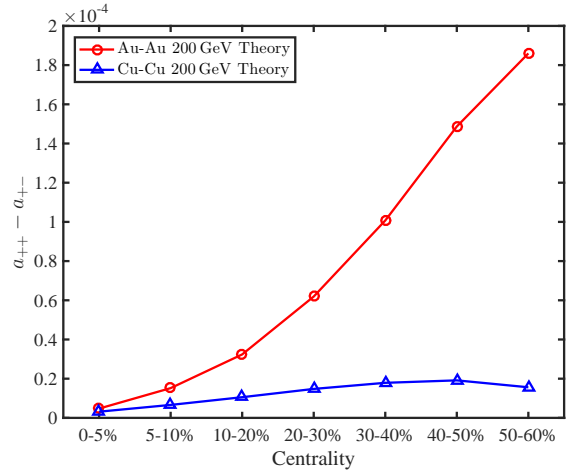}
\caption{\label{fig5}The estimation of the correlator $a_{++} - a_{+-}$ for Cu-Cu collisions at $\sqrt{s} = 200\,\mathrm{GeV}$.}
\end{figure}

The experimental data of $\delta$ correlator for Cu-Cu collisions is absent, so we can not get its background-subtracted correlator $H$. Therefore, we estimated the correlator $a_{++} - a_{+-}$ for Cu-Cu collisions at $\sqrt{s} = 200\,\mathrm{GeV}$ using the same parameter settings from Au-Au collisions. The results are plotted in Fig.~\ref{fig5}. The result of Au-Au collisions at $\sqrt{s} = 200\,\mathrm{GeV}$ is also plotted in Fig.~\ref{fig5} for comparison. As can be seen from the figure, the correlator of Au-Au collisions is much larger than that of Cu-Cu collisions.

The main reason for this result is that, at the same centrality, the smaller the system size, the smaller the initial magnetic field. To illustrate this point, we present the centrality dependence of $Q_s^2$, $t_0$, and $eB_y^0$ for Cu-Cu collisions at $\sqrt{s} = 200\,\mathrm{GeV}$ in Tab.~\ref{tab:CuRHICdata}. Comparing Tab.~\ref{tab:RHICdata} and Tab.~\ref{tab:CuRHICdata}, we find that the initial magnetic field of Cu-Cu collisions is 3 to 5 times lower than that of Au-Au collisions.

\begin{table}
\centering
\caption{\label{tab:CuRHICdata}Centrality dependence  of $Q_s^2$, $t_0$ and $eB_y^0$ for Cu-Cu collisions at $\sqrt{s} = 200\,\mathrm{GeV}$.}
\begin{tabular}{lllll}
\hline
  Centrality & $b$ & $Q_s^2$ & $t_0$ & $eB_y^0$ \\
             & ($\mathrm{fm}$)  &  ($\mathrm{GeV}^2$) & ($\mathrm{fm}$) & ($\mathrm{MeV}^2$) \\
\hline
0--5\% & 1.75 & 1.62 & 0.155 & 619.7 \\
5--10\% & 2.80 & 1.55 & 0.159 & 782.5 \\
10--20\% & 3.97 & 1.43 & 0.165 & 818.5 \\
20--30\% & 5.15 & 1.26 & 0.176 & 728.9 \\
30--40\% & 6.10 & 1.07 & 0.191 & 590.5 \\
40--50\% & 6.92 & 0.88 & 0.211 & 446.0 \\
50--60\% & 7.68 & 0.66 & 0.242 & 311.4 \\
  \hline
\end{tabular}
\end{table}

\section{Summary}\label{summary}
In this paper, we estimate the CME in relativistic heavy-ion collisions considering the magnetic field response of the QGP medium. The QGP medium has a significant influence on the time evolution of the magnetic field. To estimate the magnetic field, we adopted the optimistic assumption that assuming the electric conductivity $\sigma$ of the medium is large enough to take QGP as an ideally conducting plasma. The time evolution of the magnetic field is substituted into the KMW model to estimate the CME in relativistic heavy-ion collisions.

We compare our calculation results with the experimental resutls of background-subtracted correlator $H$. The results show that our method explains the experimental data better at RHIC than at LHC. The failure of our method at LHC may be due to the assumption that the magnetic field is homogeneously distributed in space which is not satisfied at LHC. The specific explanation remains to be further studied. The centrality dependence of correlator $|a_{+-}|/a_{++}$ for different screening length is presented, and the results are similar to that of Ref.~\cite{Kharzeev:2007jp}. At last, we give an estimation of correlator $a_{++} - a_{+-}$ for Cu-Cu collisions at RHIC energy, and find it is much smaller than that of Au-Au collisions with same energy and centrality.

\section*{Acknowledgments}
This work was supported by National Natural Science Foundation of China (Grant Nos. 11747115, 11475068), the CCNU-QLPL Innovation Fund (Grant No. QLPL2016P01), the Excellent Youth Foundation of Hubei Scientific Committee (Grant No. 2006ABB036).

\section*{References}

\bibliography{ref}

\end{document}